\documentclass[aps,prl,unsortedaddress,superscriptaddress,a4paper,nofootinbib,nobalancelastpage,showpacs,twocolumn]{revtex4-1}
\usepackage{latexsym,amsbsy}
\usepackage{multirow}
\usepackage[dvips]{graphicx}

\begin{document}
\baselineskip=11.4pt
\title{Reply to "Comment on Strangeness -2 hypertriton"} 

\author{H.~Garcilazo} 
\affiliation{Escuela Superior de F\' \i sica y Matem\'aticas (IPN), 07738 M\'exico D.F., Mexico} 

\author{A.~Valcarce}  
\affiliation{Departamento de F\'\i sica Fundamental,
Universidad de Salamanca, E-37008 Salamanca, Spain}
 
\maketitle 
In this Reply we argue that the conclusions derived in
Ref.~\cite{gg} are questionable.
In Ref.~\cite{HV} we reported the following novelties:
{\bf 1)} For the first time the Faddeev equations for 
the coupled $\Lambda\Lambda N-\Xi NN$
system have been solved. {\bf 2)} For the first time the previous formalism has been
applied to the three-baryon strangeness $-2$
system with a single model for the interactions
of the two-body subsystems.
{\bf 3)} For this model, the $\Lambda\Lambda N$ system alone does not 
present a bound state, but the three-body system with quantum numbers 
$(I,J^P)=(\frac{1}{2},\frac{1}{2}^+)$ is slightly below threshold.
  
Ref.~\cite{gg} has taken alone the uncoupled $\Lambda\Lambda$ scattering
length of the model of Ref.~\cite{HV} (that we provided to the author), and 
has compared with results of three-body calculations of the $\Lambda\Lambda\alpha$ system in which either
unrealistic separable potentials have been used for the two-body subsystems
(Ref. [7] of the Comment) or the coupling $\Lambda\Lambda-N\Xi$ has been 
included only in an effective manner (Ref. [6] of the Comment).
From this comparison Ref.~\cite{gg} speculates about the results of the model
of Ref.~\cite{HV} for the binding energy of the ${}_{\Lambda\Lambda}^6$He.
The binding energy obtained is attributed to the single 
piece picked up from Ref.~\cite{HV}.

The failure of the reasoning of Ref.~\cite{gg} is demonstrated in the following table, where 
the binding energy of the strangeness $-2$ hypertriton ($B_{\hat S=-2}$)
measured with respect to the $NH$ threshold has been calculated for one of the models
of Table III of Ref.~\cite{HV} ($a^{N\Lambda}_{1/2,1}=-1.58$ fm,
$a^{N\Lambda}_{1/2,0}=-2.48$ fm, signs are changed to use the convention of the comment) for different values of the 
uncoupled $\Lambda\Lambda$ scattering length but which describe
equally well the available experimental data.
\begin{table}[h!] 
\begin{ruledtabular} 
\begin{tabular}{ccc}
 $-a_{\Lambda\Lambda}$ (fm) & $B_H$ (MeV)& $B_{\hat S=-2}$ (MeV) \\ \hline
 3.3  &  6.928   &   0.577 \\
 2.3  &  6.191   &   0.640 \\
 1.3  &  4.962   &   0.753 \\
 0.5  &  3.250   &   0.927 \\
\end{tabular}
\end{ruledtabular} 
\label{t2}
\end{table}
These results rule out the arguments of Ref.~\cite{gg} as we
show next. Ref.~\cite{gg} argues that the H dibaryon and the strangeness $-2$ 
hypertriton are both bound because the CCQM generates a $\Lambda\Lambda$
uncoupled scattering length of $-$3.3 fm and therefore since in 
${}_{\Lambda\Lambda}^6$He only the uncoupled $\Lambda\Lambda$
scattering length acts, due to the Pauli principle, this model would
lead to a very large ${}_{\Lambda\Lambda}^6$He binding energy which contradicts the
experiment. However, as shown in the previous table, the existence of both a 
bound H dibaryon and a bound strangeness $-2$ 
hypertriton is compatible with a small 
$\Lambda\Lambda$ uncoupled scattering length which kills the
argument of Ref.~\cite{gg}. The Pauli principle acts strongly in 
${}_{\Lambda\Lambda}^6$He because there is no room for more that
four nucleons in $S$ wave while in
${}_{\Lambda\Lambda}^3$H the full $N\Xi$ interaction can act in
$S$ wave. Thus, one cannot say that our $YY$-interaction model
overbinds the ${}_{\Lambda\Lambda}^6$He until a calculation of that 
system using the model of Ref.~\cite{HV} has been done.

The procedure of Ref.~\cite{gg} contains other uncertainties that makes
any final conclusion doubtful. Ref.~\cite{Nem03} warned about 
the use of $NN$, $N\Lambda$ and $\Lambda\Lambda$ two-body 
interactions improved for the description of the ${}_{\Lambda\Lambda}^6$He to 
study other double $\Lambda$ hypernuclei, as for example the ${}_{\Lambda\Lambda}^4$He. 
They demonstrate that a choice of the $N\Lambda$ interaction different 
to the references used in Ref.~\cite{gg} gives binding for the ${}_{\Lambda\Lambda}^4$He~\cite{Nem05}
for a wide range of $\Lambda\Lambda$ scattering lengths~\cite{Nem03}.
This state would be unbound for the prescriptions used in Ref.~\cite{gg}.
Refs.~\cite{Nem03,Him06} also called the attention about the 
$\alpha\Lambda\Lambda$ three-body model used in Ref.~\cite{gg}, that
might be inappropriate for deducing the $\Lambda\Lambda$
interaction in free space from the experimental information
on $B_{LL}({}_{\Lambda\Lambda}^6$He). All these details
are circumvented in Ref.~\cite{gg}.

Ref.~\cite{gg} writes that "the latest HAL QCD lattice-simulation analysis 
locates the H dibaryon near the $\Xi N$ threshold.", quoting Ref.~\cite{Ino12}.
Immediately after this sentence one can read in Ref.~\cite{Ino12} "This is however
not a final conclusion due to various approximations about the 
SU(3) breaking ... currently underway lattice QCD simulations ... will eventually
clarify the nature of the elusive H-dibaryon".
Quadratic and linear extrapolations to the physical point, 
not performed in Ref.~\cite{Ino12}, using the results of the
HAL QCD and NPLQCD collaborations have been presented in Ref.~\cite{Bea11},
allowing in both instances for a bound H-dibaryon or a near-threshold 
scattering state. This illustrates the actual uncertainties about the H dibaryon.

In summary, for all these reasons the 
conclusions of Ref.~\cite{gg} are questionable.

\end{document}